# High Performance Multiple Sequence Alignment Algorithms for Comparison of Microbial Genomes

Manal Helal,[1,2]　　Hossam El-Gindy,[2]　　Bruno Gaeta,[2]　　Vitali Sinchenko[1]

`mhelal@usyd.edu.au;`　`hossam@cse.unsw.edu.au;`　`bgaeta@unsw.edu.au;`　`vsintchenko@usyd.edu.au`

[1] Centre for Infectious Diseases and Microbiology, Western Clinical School, University of Sydney, Sydney, Australia
[2] School of Computer Science and Engineering, University of New South Wales, Sydney, Australia

**Abstract**
Advances in gene sequencing have enabled *in silico* analyses of microbial genomes and have led to the revision of concepts of microbial taxonomy and evolution. We explore deficiencies in existing multiple sequence global alignment algorithms and introduce a new indexing scheme to partition the dynamic programming algorithm hypercube scoring tensor over processors based on the dependency between partitions to be scored in parallel. The performance of algorithms is compared in the study of *rpoB* gene sequences of *Mycoplasma* species.

## 1 Introduction

Accurate sequence alignments are essential for sequence comparison and for building structure-activity models. High performance multiple sequence alignment (MSA) algorithms have been suggested to improve the validity and relevance of *in silico* genomic experiments. Existing MSA heuristics introduce significant biases affecting the quality of alignments [1] as they apply progressive pair wise alignments of each pair of sequences using different tree building methods. Advances in gene sequencing have enabled large-scale analyses of microbial genomes and have led to the revision of concepts of taxonomy and evolution. Several genes have been suggested as discriminatory targets. The *rpoB* gene, or the DNA-dependent RNA polymerase gene, has been proposed as a genome similarity predictor and an alternative to 16S rDNA gene sequencing for biodiversity studies. This study aimed to optimize and parallelise MSA by applying an innovative indexing scheme with search space partitioning to the study of *rpoB* gene sequences.

## 2 Method and Results

**Search space partitioning:** The mathematics of arrays and the PSI-Calculus [2] have made the high dimensional hyperplane partitioning methods invariant of shape and dimension feasible. They allow for the representation of a hyperplane as an exponentially growing tree from one node at the first level till the middle level then decreasing in reverse symmetry to one node at the last level. Each node in the tree becomes a vector of indices representing the partition-starting index (a multiple of the partition size used). All nodes on the same level can run in parallel in one wave of computation if they are divided over the available processors. Then, communication of overlapping cells takes place, and another wave of computation continues until the end of the tree. The division over the processors at each wave takes care of dependency across waves to keep communication cost down to the minimum. The exponential growth is shifted to employment of multiple processors, keeping the partition size reasonable to the cache of the processors used.

**DNA sequences:** Nucleotide sequences of *rpoB* gene from *Mycoplasma* species were chosen as the validation set as *Mycoplasma* is the simplest free-living organism with the minimal genome. The GenBank accession numbers used in this study were AY191418, AY191419, AY191420, AY191421, AY191423, AY191424, AY191425, AY191433, and AY191437. Sequences were aligned using the proposed mmDst algorithm, and existing methods such as clustalW, Muscle and Tcoffee. The algorithm was tested on a SunFire X2200 machine with 2xAMD Opteron quad processors of 2.3 GHz, 512 Kb L2 cache and 2 MB L3 cache on each processor, and 8GB RAM. The sequences were aligned in full search space (scoring all partitions) using 8 processors and partition size of 20. The dynamic programming scoring defined for two and three dimensions was extended too arbitrary dimension.

**Alignment scoring:** The score of one cell in the hyperplane was based on the maximum values of the $2^k$-1 neighbours' temporary scores. The latter was calculated as the pair wise scores of all its corresponding residues on all dimensions (sequences) corresponding to a decremented index element from the current cell index to the neighbour index, plus multiplication of the gap score by the number of un-decremented index elements. The following penalties were applied: flat gap penalty = -2, mismatch score = 0, and match score = 2 in contrast to affine gap (minimum 0.05 for gap extension) for ClustalW, and gap opening penalty of 1 and gap penalty of 0 employed by Muscle and Tcoffee, respectively. Using a simple sum of pairs score with the above scoring scheme, mmDst consistently produced alignments with higher scores than clustalW and Tcoffee. The scores were 5003, 6233, 6467 and 6517 for clustalW, Tcoffee, mmDst and Muscle, respectively. Table 1 reports the weights for sequences from the sequence alignments using different methods as calculated by MASH (http://timpani.bmr.kyushu-u.ac.jp/~mash/weight_ex.html).

**Table 1: Sequence Weights from the different alignments (Overall Std. Dev.: 0.028351)**

|  | Difference Method | | | | | Henikoff-Henikoff method | | | | |
|---|---|---|---|---|---|---|---|---|---|---|
|  | MmDst | ClustalW | Tcoffee | Muscle | Std. Dev. | MmDst | ClustalW | Tcoffee | Muscle | Std. Dev. |
| M.arthritidis | 0.098665 | 0.105587 | 0.101317 | 0.100486 | 0.002933 | 0.084044 | 0.089908 | 0.091009 | 0.089298 | 0.003096 |
| M.bovis | 0.106124 | 0.107244 | 0.107623 | 0.105600 | 0.000945 | 0.083984 | 0.097180 | 0.094332 | 0.092975 | 0.005699 |
| M.buccale | 0.093038 | 0.090909 | 0.092209 | 0.092304 | 0.000885 | 0.069931 | 0.084373 | 0.080028 | 0.079753 | 0.006105 |
| M.fermentans | 0.105862 | 0.108665 | 0.110706 | 0.107262 | 0.002067 | 0.087820 | 0.102429 | 0.097179 | 0.095294 | 0.006048 |
| M.gallisepticum | 0.132819 | 0.125947 | 0.134529 | 0.134620 | 0.004106 | 0.170381 | 0.168571 | 0.175023 | 0.186921 | 0.008258 |
| M.genitalium | 0.130463 | 0.126894 | 0.127663 | 0.129890 | 0.001718 | 0.151906 | 0.135856 | 0.141880 | 0.139587 | 0.006864 |
| M.hominis | 0.100628 | 0.103456 | 0.098374 | 0.098057 | 0.002496 | 0.079683 | 0.090872 | 0.087999 | 0.084857 | 0.004791 |
| M.pneumoniae | 0.136613 | 0.132576 | 0.133268 | 0.138456 | 0.002782 | 0.199205 | 0.143638 | 0.149545 | 0.152684 | 0.025568 |
| M.salivarium | 0.095786 | 0.098722 | 0.094311 | 0.093327 | 0.002352 | 0.073045 | 0.087172 | 0.083005 | 0.078630 | 0.006051 |

**Figure 1: Dendrograms produced by mmDst (a), clustalW (b), Tcoffee (c), and Muscle (d).**

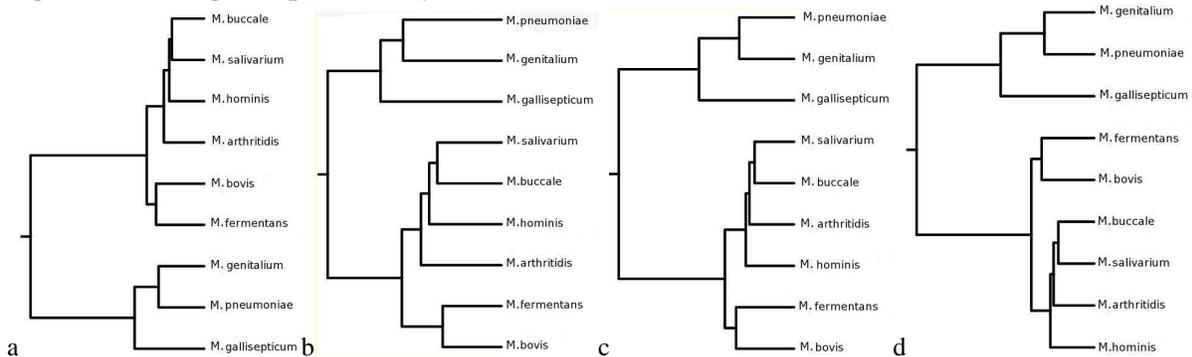

## 3 Discussion

Optimized MSA algorithm based on an innovative indexing scheme with search space partitioning produced consistently better quality alignments than clustalW, and comparable to Muscle and Tcoffee in the study of *rpoB* gene sequences of *Mycoplasma* species. The algorithm can be re-configured to different high performance computing architectures by adjusting parameters such as: partition size, search space reduction factor, and number of CPUs. We are currently developing a method for automatic selection of optimal parameters given the constraints of the architecture and the data set. High performance MSA methods can optimise alignments of nucleotide or amino acid sequences for comparative genomics of infectious diseases, for the identification of infectious pathogens based on gene sequencing and for sequencing-based analysis of microbial virulence and drug resistance.